# THE IMPERATIVES OF COSMIC BIOLOGY

Carl H. Gibson [1,2]

[1] University of California San Diego, La Jolla, CA 92093-0411, USA
[2] cgibson@ucsd.edu, http://sdcc3.ucsd.edu/~ir118

N. Chandra Wickramasinghe[3,4]

[3] Cardiff Centre for Astrobiology, Cardiff University, 2 North Road, Cardiff CF10 3DY, UK
[4] ncwick@googlemail.com, http://www.astrobiology.cf.ac.uk

## Abstract

The transformation of organic molecules into the simplest self-replicating living system – a microorganism – is accomplished from a unique event or rare events that occurred early in the Universe. The subsequent dispersal on cosmic scales and evolution of life is guaranteed, being determined by well-understood processes of physics and biology. Entire galaxies and clusters of galaxies can be considered as connected biospheres, with lateral gene transfers, as initially theorized by Joseph (2000), providing for genetic mixing and Darwinian evolution on a cosmic scale. Big bang cosmology modified by modern fluid mechanics suggests the beginning and wide intergalactic dispersal of life occurred immediately after the end of the plasma epoch when the gas of protogalaxies in clusters fragmented into clumps of planets. Stars are born from binary mergers of such planets within such clumps. When stars devour their surrounding planets to excess they explode, distributing necessary fertilizing chemicals created only in stars with panspermial templates created only in adjacent planets, moons and comets, to be gravitationally collected by the planets and further converted to living organisms. Recent infrared images of nearby star forming regions suggest that life formation on planets like Earth is possible, but not inevitable.

## 1. Introduction

The contemporary scientific approach to the problem of the origin of life is being shaped mainly within the emergent discipline of astrobiology – a discipline that combines the sciences of astronomy, space science and biology. The fact that water and complex carbon-based organic molecules are ubiquitously present outside the Earth is leading some scientists towards a possibly erroneous point of view: that life is easily generated *in situ* from non-living matter – the ancient doctrine of spontaneous generation being essentially revived.

The astronomical origin of the "stuff" of life at the level of atoms is beyond dispute. The chemical elements C,N,O,P… and the metals that are present in all living systems were synthesised from the most common element hydrogen in nuclear reactions that take place in the deep interiors of stars (Burbidge, Burbidge, Fowler and Hoyle, 1957). The explosions of supernovae scatter these atoms into the frozen protoglobularstarcluster clumps (PGCs) of primordial earth-mass planets (PFPs) from which new stars, larger planets, moons and comets are produced by chains of PFP mergers (Gibson 2009ab, Schild and Gibson 2010, Gibson and Schild 2010; Joseph and Schild 2010). The dark matter of galaxies was independently discovered to be planets in clumps (Gibson 1996, Schild 1996) by fluid mechanical theory and by quasar microlensing, respectively. The combination of atoms into organic molecules can proceed in interstellar planets via well-attested chemical pathways, but only to a certain limiting level of complexity. The discovery of biochemical molecules in space, including comets and meteorites, crosses a limiting threshold, although the precise level of biochemical complexity that *can* be reached through chemistry alone is still in dispute (Joseph 2009a). New infrared space telescopes offer hope that the probability estimates for life on planets can be refined. All galaxies appear to support life (Joseph and Wickramasinghe 2010a), but not all PGCs.

The only secure empirical fact relating to the origin of life is encapsulated in a dictum eloquently enunciated by Louis Pasteur *Omne vivum e vivo*—all life from antecedent life (1857). If life is always derived from antecedent life in a causal chain that is clearly manifest in present day life and through the fossil record, the question naturally arises as to when and where this connection may have ceased. The continuation of the life-from-life chain to a time before the first life appears on our planet and before the Earth itself formed implies the operation of "panspermia". The basic concept of panspermia has an ancient history going back centuries - to the time of classical Greece and even before – referring in general to the widespread dispersal of the "seeds of life" in the cosmos (Arrhenius, 1908; Hoyle and Wickramasinghe, 2000). Critics of panspermia often say that such theories are of limited value because they do not address the fundamental question of origins. Nevertheless the question of whether life originated *in situ* on Earth, or was delivered here from the wider universe constitutes a scientifically valid line of inquiry that needs to be pursued.

Whilst the Francis Crick and Leslie Orgel's idea of directed panspermia transfers the problem of origin to another site, possibly invoking intelligent intervention (Crick and Orgel, 1973), Fred Hoyle and one of the present authors (Wickramasinghe et al., 2009) have attempted to expand the domain in which cosmic abiogenesis *may* have occurred, focussing in particular on the totality of comets in our galaxy. Like Crick and Orgel (1973) Hoyle and Wickramasinghe were influenced by the estimated super-astronomical odds against the transition from organic molecules to even the most primitive living system (Hoyle and Wickramasinghe, 1982).

The currently fashionable view that all extraterrestrial organics arise abiotically – that is to say through non-biologic processes – has no secure empirical basis and is likely to be flawed (Joseph 2009a). On the Earth it is clear that life processes account for almost all the organic molecules on the planet. If biology can somehow be shown to be widespread on a cosmic scale, the detritus of living cells would also be expected to be widely distributed in the Cosmos. The bulk of the organic molecules in space would then be explained as break-up products of life-molecules. Inorganic processes can scarcely be expected to compete with biology in the ability to synthesise systems of biochemicals resembling the detritus of biology. So wherever complex organics are found in an astronomical setting, one might legitimately infer that biology has spread astrophysically (Wickramasinghe, 2010).

## 2. Abiogenesis

But what then of a first origin of life?

Charles Darwin, who laid the foundations of evolutionary biology, never once alluded to the origin of life in his 1859 book On the Origin of Species (Darwin, 1859). He had, however, thought about the problem and formulated his own tentative position in a letter to Joseph Hooker in 1871 thus:

> "….It is often said that all the conditions for the first production of a living organism are now present, which could ever have been present. But if (and oh! what a big if!) we could conceive in some warm little pond, with all sorts of ammonia and phosphoric salts, light, heat, electricity, &c., present, that a proteine compound was chemically formed ready to undergo still more complex changes, at the present day such matter would be instantly absorbed, which would not have been the case before living creatures were found."

Darwin's prescient remarks provided the basic scientific framework for exploring the problem of abiogenesis throughout the 20th century and beyond. In the late 1920's A.I. Oparin (1953) and J.B.S. Haldane (1929) fleshed out Darwin's thoughts into the familiar "Primordial Soup Theory", proposing that the atmosphere of the primitive Earth comprised of a reducing mixture of hydrogen, methane and ammonia and other compounds from which the monomers of life could be readily generated. Primitive 'lightening' and solar ultraviolet provided the energy to dissociate these molecules, and the radicals so formed recombined through a cascade of chemical reactions to yield biochemical monomers such as amino acids, nucleotide bases and sugars. The classic experiments of Stanley Miller and Harold Urey (1959) demonstrated the feasibility of the chemical processes proposed by Oparin and Haldane, and this led to the belief that life could be generated *de novo* as soon as the biochemical monomers were in place. The formation of the first fully-functioning, self-replicating life system with the potential for Darwinian evolution is riddled with the difficulty of beating super-astronomical odds and still remains an elusive concept.

## 3. Emergence of the idea of a cosmic imperative

Arguments relating to the *improbability* of life emerging from non-life has had a chequered history. The image of a "watchmaker" and of the complexity of a pocket watch was infamously introduced by William Paley, (1743–1805):

> "inspect the watch, we perceive . . . that its several parts are framed and put together for a purpose, e.g. that they are so formed and adjusted as to produce motion, and that motion so regulated as to point out the hour of the day; that if the different parts had been differently shaped from what they are, or placed after any other manner or in any other order than that in which they are placed, either no motion at all would have been carried on in the machine, or none which would have answered the use that is now served by it..."

The inference that the Watchmaker was God was of course unnecessary, erroneous and scientifically inadmissible.

Nobel Laureate Jacques Monod, who made fundamental contributions to enzymology and gene expression, revived the same argument in a biochemical context and advanced the view that life on Earth arose by a freak combination of exceedingly improbable circumstances, so improbable indeed that it may have arisen only once in the entire history of the universe. He wrote in his book Chance and Necessity (Monod, 1971):

> "Man at last knows he is alone in the unfeeling immensity of the universe, out of which he has emerged only by chance…"

Monod's essentially geocentric view was unpalatable to many and a majority of biologists opted to follow yet another Nobel Laureate Christian de Duve, who championed the idea that life must be a *cosmic imperative* (Duve, 1996). His ideas gained fashion because astrophysicists were discovering organic molecules in interstellar dust clouds. But does the widespread occurrence of these molecules provide a means of beating the incredible odds of getting the correct arrangements that lead to the emergence of life everywhere?

The improbability argument stated here has its critics of course. But the case for an infinitesimally small probability of abiogenesis cannot be peremptorily dismissed as wrong: at worst it might be considered *indeterminate*. Scientific arguments for abiogenesis remain ill-defined to the extent that no processes have been identified that can lead from prebiotic chemistry to life. The enormous probability hurdle needs somehow to be overcome.

The *a priori* probability for the emergence of crucial molecular arrangements needed for life (e.g. in the enzymes) have been variously estimated as being about 1 in $10^x$, where x is in the range ~ 100 to 40,000 (Orgel and Crick, 1975; Hoyle and Wickramasinghe, 1982). Even assuming the lower value in this range the probability of abiogenesis leading to life remains truly minuscule.

Since we know that life exists on the Earth two possibilities remain open. It arose on Earth against incredible odds, in which case the emergence of life on other planets would be multiply improbable, and an argument could then be made for life being unique to the Earth or to this solar system – a position that was passionately espoused by Jacques Monod. Such a pre-Copernican viewpoint is less appealing than the idea that life emerged on a cosmic scale that transcended the scale of the solar system, even the scale of the entire galaxy.

The resistance to all such ideas, in our view, is due to a fear that the argument would give succour to creationism. This is unfounded of course because *abiogenesis* must surely have happened according to our view – but only as an exceedingly rare, even unique event in the history of the cosmos. The spread and diffusion of life even from a single event of origination then turns out to be essentially unstoppable (Joseph and Wickramasinghe 2010a).

The probability of *abiogenesis* is greatly increased by recognizing that the interstellar medium is dominated by planets in clumps merging to make stars (Gibson 2009ab). Rather than the $10^{12}$ non-merging planets expected per galaxy in old cosmology, new cosmology requires more than $10^{19}$ and their moons and comets constantly recycling life giving stardust from the stars they produce and destroy.

## 4. Interstellar organics molecules and the origin of life on the Earth

In recent years Astrobiology has taken up the challenge of extending the Oparin-Haldane ideas of abiogenesis to a wider cosmic canvas. This has been prompted in large measure by the discovery of biochemically relevant molecules such as polycyclic aromatic hydrocarbons (PAHs: *ie*, oil-like) in interstellar space, an identification first reported in the journal *Nature* by Fred Hoyle and one of us in 1977 (Hoyle and Wickramasinghe, 1977, 2000). Such molecules have now been inferred to exist in vast quantity not only within the Milky Way but in external galaxies as well (Wickramasinghe et al., 2004).

Figure 1 shows a Herschel infrared telescope image of the Aquila Nebula which contains clouds choc-a-bloc with organic molecules including PAHs and is an active site of star-births, the youngest stars being younger than a few million years. Such stellar and planetary nurseries are considered by many to be regions where a Urey-Miller-type chemistry is taking place on a grand scale. An unproved assertion is that life must be the inevitable end result of such chemistry- a cosmic imperative.

We argue as an alternative that such interstellar clouds may rather represent not only a nursery for stars but a graveyard of life – polycyclic aromatic hydrocarbons and other organic molecules present here arise from the destruction and degradation of life.

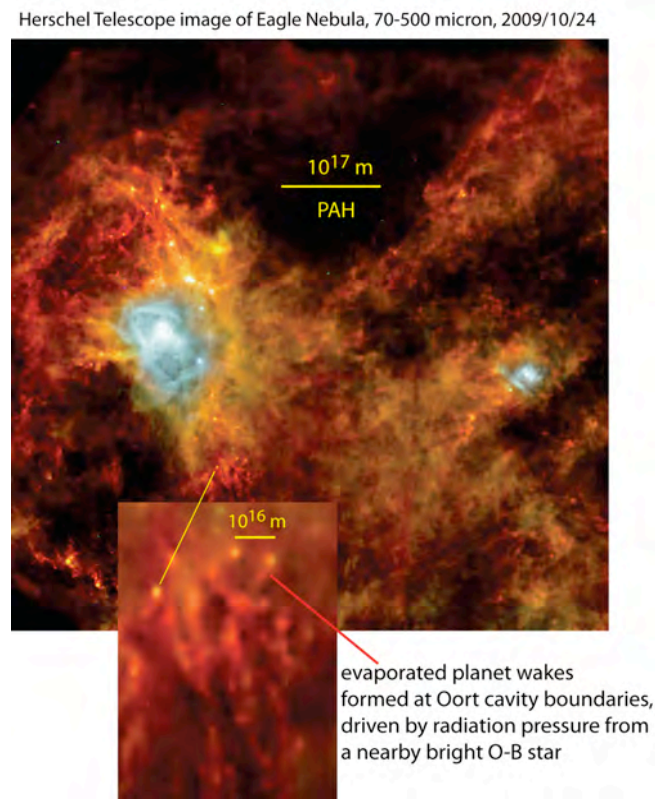

Fig. 1 The Aquila Nebula – A dark cloud at the heart of the Eagle Nebula photographed by the Herschel Space Observatory - a stellar nursery and host for life. The distance is $10^{19}$ m from Earth, so PGCs have been reduced in size by the stratified turbulence of the Milky Way spiral accretion disk. Despite the enhanced

mixing, only a fraction of the planet clumps show PAH signs of life. The dust wakes of ~ 1% $M_{SUN}$ brown dwarfs in the insert reflect O-B erosion of ambient planets outside $10^{15}$ m Oort cavities (Fig. 2).

The Aquila Nebula is only about $10^{19}$ m from Earth, embedded in the stratified turbulence of the spiral arms of the Milky Way. Over the 13.7 Gyr since their creation, PGCs have frozen and evaporated from the primordial $10^{20}$ m (Nomura scale) central core of the galaxy to form a kpc ($3\times10^{19}$ m) thick accretion disk within the $10^{22}$ m diameter dark matter galaxy halo of mostly lifeless and metal free PGCs. Even in this relatively well mixed and fertilized region of the galaxy, only a small fraction of the image shows PAH signs of life, consistent with the Hoyle-Wickramasinghe 1972 etc. claims that the odds for *abiogenesis* as the source of life are small, certainly on such an insignificant planet as the Earth.

Support for the idea that life originated on Earth in a primordial soup is beginning to wear thin in the light of recent geological and astronomical evidence. It is becoming clear that life arose on Earth almost at the very first moment that it could have survived. During the period from about 4.3-3.8 Gyr ago (the Hadean Epoch) the Earth suffered an episode of heavy bombardment by comets and asteroids. Rocks dating back to the tail end of this epoch reveal evidence of an excess of the lighter isotope $^{12}C$ compared with $^{13}C$ pointing to the action of microorganisms that preferentially take up the lighter isotope from the environment (Mojzsis et al, 1996; Manning et al, 2006).

**5. Origin of life in comets**

If we reject the option of an origin of life on the Earth against incredible odds as discussed in section 3, a very much bigger system and a longer timescale was involved in an initial origination event, after which life was transferred to Earth.

The $4\times10^{17}$ m PGC molecular clouds of planets and their atmospheres in the galaxy are of course much bigger in scale than anything on Earth, but in the constantly merging and recycling planetary interstellar medium one can achieve the production of organic molecules through a variety of liquid and gas-phase chemistry, particularly if supernovae in the planet clump have provided sufficient fertilizer and seeds. These organic molecules must enter a watery medium in suitably high concentrations to begin the presumptive prebiotic chemistry that eventually leads to life. In the formation of a solar system (and planetary nebulae systems approaching supernovae such as the Helix seen in Fig. 2), numerous solid objects form, such as larger planets, moons and comets. These originally icy objects would contain the molecules of the parent interstellar clouds of planets, and for a few million years after they condensed would have liquid water interiors due to the heating effect of radioactive decays (J. Wickramasinghe et al, 2009). If microbial life were already present in the parent interstellar medium, the newly formed comets could serve to amplify it.

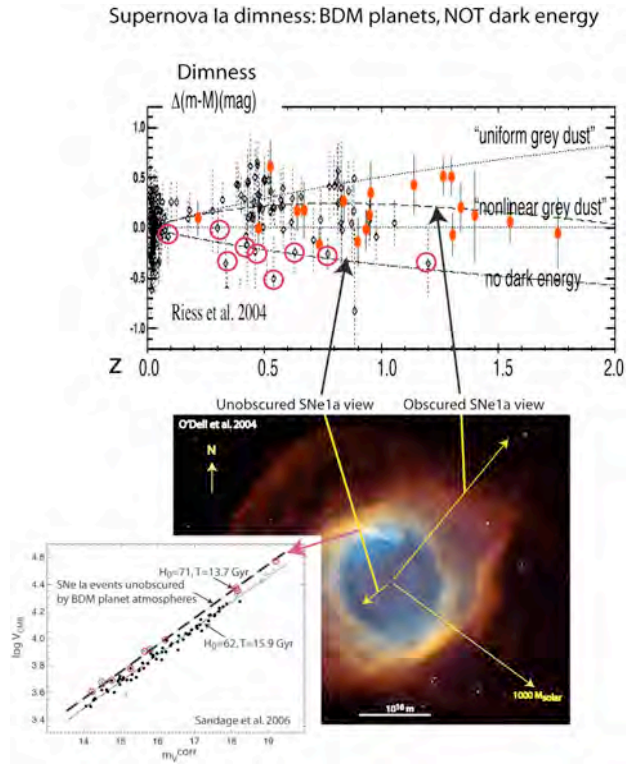

Fig. 2  Helix, the nearest planetary nebula to Earth (Gibson 2009b, Fig. 8, p 58).  Dimness of supernovae Ia events is the result of ambient planet atmospheres on some lines of sight (top), not cosmological constant Λ dark energy (antigravity) accelerating the expansion of the universe.  Similar systematic dimming errors lead to overestimates of the age of the universe (lower left).  The spinning central white dwarf reflects overfeeding by planets and comets coming from the edge of the $10^6$ m Oort cavity within the PGC planet clump.

Figure 2 shows the Helix planetary nebula and its many thousands of ambient planets revealed in the optical frequency band by the Hubble Space Telescope.  The central white dwarf star is compressed by accreted baryonic dark matter (BDM) planets.  To conserve angular momentum the shrinking white dwarf rapidly spins, powering a plasma jet from the accreted materials.  The plasma jet and equatorial accretion disk radiation evaporates frozen H and $^4$He ambient planets to maximum scales of $10^{13}$ m for Earths, and to $3\times10^{13}$ m for Jupiters, as observed in photoionization images.  When the star explodes its maximum brightness is the proper indication of its distance, not lines of sight dimmed by atmospheres of evaporated PFPs outside the Oort cavity.

Ignorance of frozen primordial planets in clumps as the dark matter of galaxies has led to a false indication of Λ dark energy by systematic SNIa dimming errors.  Antigravity forces of the big bang are supplied by vortex dynamics of big bang turbulence, not a cosmological constant Λ.  $Λ_{turb}$ antigravity (turbulent dark energy) powers the initial expansion of the fireball of Planck particles and antiparticles by inertial vortex forces but frictionally dissipates soon after the big bang event.

Prior to life being generated anywhere, primordial comets could provide trillions of "warm little ponds" replete with water, organics and nutrients, their huge numbers diminishing vastly the improbability hurdle for life to originate.  Recent studies of comet Tempel 1 (Figure 3) have shown evidence of organic molecules, clay particles as well as liquid water, providing an ideal setting for the operation of the "clay theory" of the origin of life (Cairns-Smith 1966; Napier et al, 2007). It can be argued that a single primordial comet of this kind will be favoured over all the shallow ponds and edges of oceans on Earth by a factor $10^4$, taking into account the total clay surface area for catalytic reactions as well as the timescale of persistence in each scenario. With $10^{11}$ comets, the factor favouring solar system comets over the totality of terrestrial "warm little ponds" weighs in at a figure of $10^{15}$, and with $10^{11}$ sun-like stars replete with comets in the entire galaxy we tot up a factor of $10^{26}$ in favour of a cometary origin life

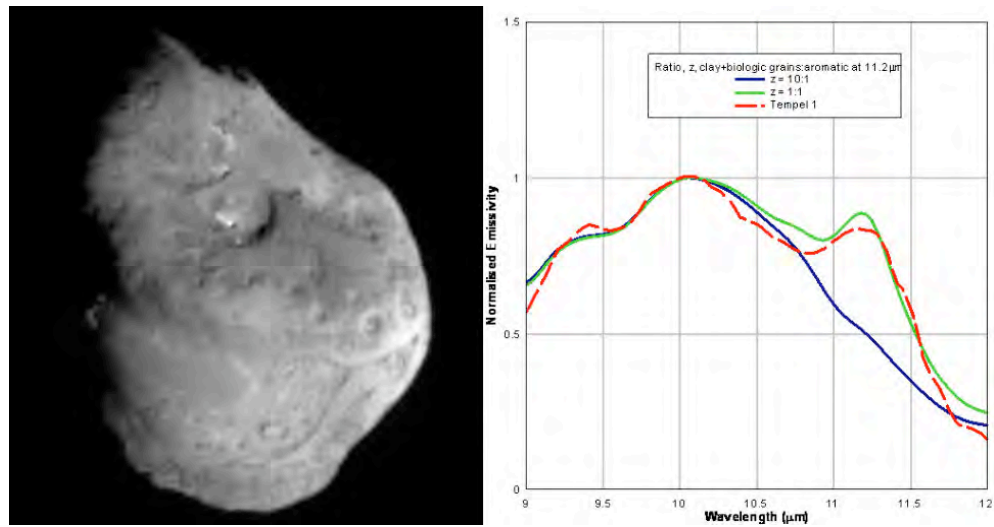

Fig. 3  Comet Tempel 1 showed evidence of relic frozen lakes and clay indicative of early contact with liquid water.

The next step in the argument is that once life got started in some comet somewhere, its spread in the cosmos becomes inevitable. Comets themselves provide ideal sites for amplification of surviving microbes that are incorporated into a nascent planetary system. Dormant microorganisms released in the dust tails of comets can be propelled by the pressure of starlight to reach interstellar clouds of planets. Transport of life in the form of microorganisms and spores within the frozen interiors of comets carries only a negligible risk of destruction, whilst transport in either naked form, within clumps of dust or within meteorites entails varying degrees of risk of inactivation by cosmic rays and UV light. It cannot be overemphasised, however, that the successful seeding of life requires only the minutest survival fraction between successive amplification sites. Of the bacterial particles included in every nascent cometary cloud only one in $10^{26}$ needs to remain viable to ensure a positive feedback loop for panspermia. All the indications are that this is indeed a modest requirement that is hard, if not impossible, to violate.

**6. Origin of Life in the Early Universe**

Whilst the distribution of life via comets is unavoidable once an origin has been achieved somewhere, the events leading to the origin itself must still remain largely speculative. The existence of life on Earth logically demands an origin, albeit an exceedingly improbable origin, somewhere in the Universe. Although the set of comets in the galaxy discussed in section 4 was seen to confer a non-trivial advantage over sites of origin on Earth by an estimated factor $10^{24}$, there would be merit in seeking even wider horizons, going back further in time in the history of the Big-Bang Universe. The earliest opportunity for such an event arises after the recombination of ions and electrons – the plasma to gas transition that took place 300,000 years ($10^{13}$ s) after the Big Bang. At that time the Universe scale of causal connection $ct$ included a baryonic mass of about $10^{48}$ kg, or $10^6$ protogalaxies, where $c$ is the speed of light and $t$ is the time since the big bang event. These starless H and $^4$He plasma clouds retained as fossils the density, strain rate and morphology of superclusters formed at $10^{12}$ s (30,000 years) and supervoids now expanded to $10^{25}$ m with $10^{24}$ m $10^{46}$ kg superclusters. Protogalaxies fragmented along weak turbulence vortex lines of the plasma epoch to form chain galaxy clusters observed by the Hubble Space Telescope ultra deep field (Gibson and Schild 2010, Fig. 1). The protogalaxy morphology is either linear or spiral at the base of the linear fragments according to Nomura direct numerical simulations of weak turbulence, giving the Nomura scale of $10^{20}$ m from the fluid mechanics. The $10^{12}$ s density $\rho \sim 10^{-17}$ kg m$^{-3}$ is the density of PGCs.

Figure 4 illustrates how cosmology modified by modern fluid mechanics (New Cosmology, Gibson 2009ab) estimates the time life should appear and how it is likely to be dispersed to cosmological scales.

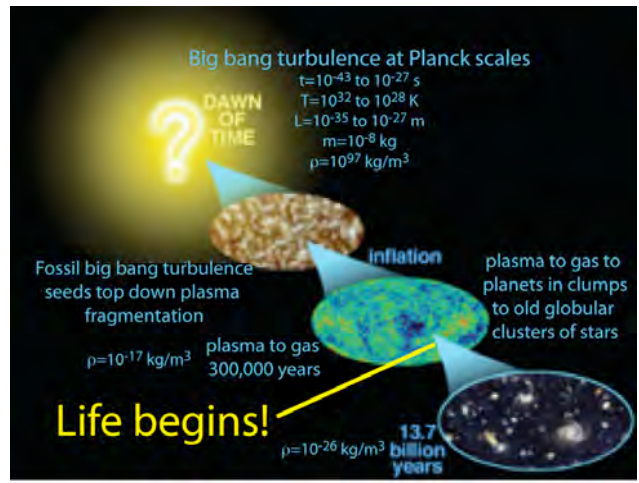

Fig 4: Depiction of Big Bang History:  Life starts about 300,000 years after the Big Bang

From the cosmic microwave background evidence the plasma as it turned to gas was incredibly quiet, reflecting its large photon viscosity and the buoyant damping of turbulence by the protosupercluster to protogalaxy mass fragmentation. It promptly (*ie*: in a first structure freefall time $\tau_{g0}$ of $10^{12}$ s) fragmented at the PGC Jeans mass of $10^{36}$ kg reflecting the sonic velocity of the gas, and the PFP Earth-mass Schwarz viscous scale reflecting a decrease of kinematic viscosity by a factor of $10^{13}$ from the photon viscosity of the plasma epoch (Gibson 1996). Old globular clusters near protogalaxy centers of gravity formed gently at first and then more violently, building a maelstrom of turbulence before collapsing to an active galactic nucleus much later. Within a few $\tau_{g0}$ time periods numerous supernovae create and distribute the chemical elements of life C, N, O, P, Fe etc. to the PGCs. Fragmentation and binary merging of planetary, moon and comet bodies in this mode of star formation continuously recycles these chemicals in the variety of watery environments needed to produce the organic monomers required for the origin of life.  Planets make stars.  Stars fertilize planets.  Comets seed them.  Life mutates from supernovae when stars overeat.

**7. Astronomical evidence**

Identifying the composition of interstellar dust in clouds such as Fig. 1 has been a high priority for astronomical research since the early 1930's (see Wickramasinghe, 1967).  The dust absorbs and scatters starlight causing extinction of the light from stars, and re-emits the absorbed radiation in the infrared.  An important clue relating to dust composition follows from studies of extinction of starlight.  The total amount of the dust has to be as large as it can be if nearly all the available carbon and oxygen is condensed into grains.  The paradigm in the 1960's that the dust was largely comprised of water-ice was quickly overturned with the advent of infrared observations showing absorptions due to CH, OH, C-O-C linkages consistent with organic polymers.  The best agreement for a range of astronomical spectra embracing a wide wavelength interval turned out to be material that is indistinguishable from freeze-dried bacteria and the best overall agreement over the entire profile of interstellar extinction was a mixture of desiccated bacteria, nanobacteria, including biologically derived aromatic (PAH) molecules as seen in Fig.5.

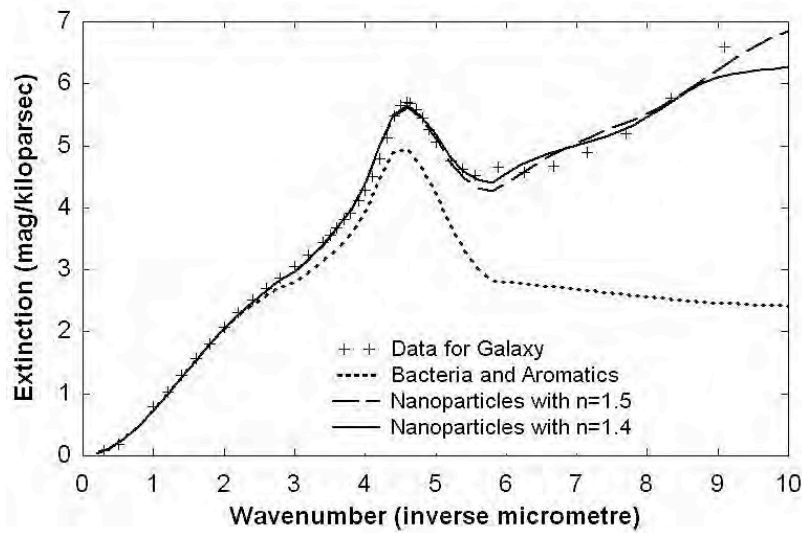

Fig.5  Agreement between interstellar extinction (plus signs) and biological models.  Mixtures of hollow bacterial grains, biological aromatic molecules and nanobacteria provide excellent fits to the astronomical data.  The 2175A hump in the extinction is caused by biological aromatic molecules (See J. Wickramasinghe et al. (2009) for details)

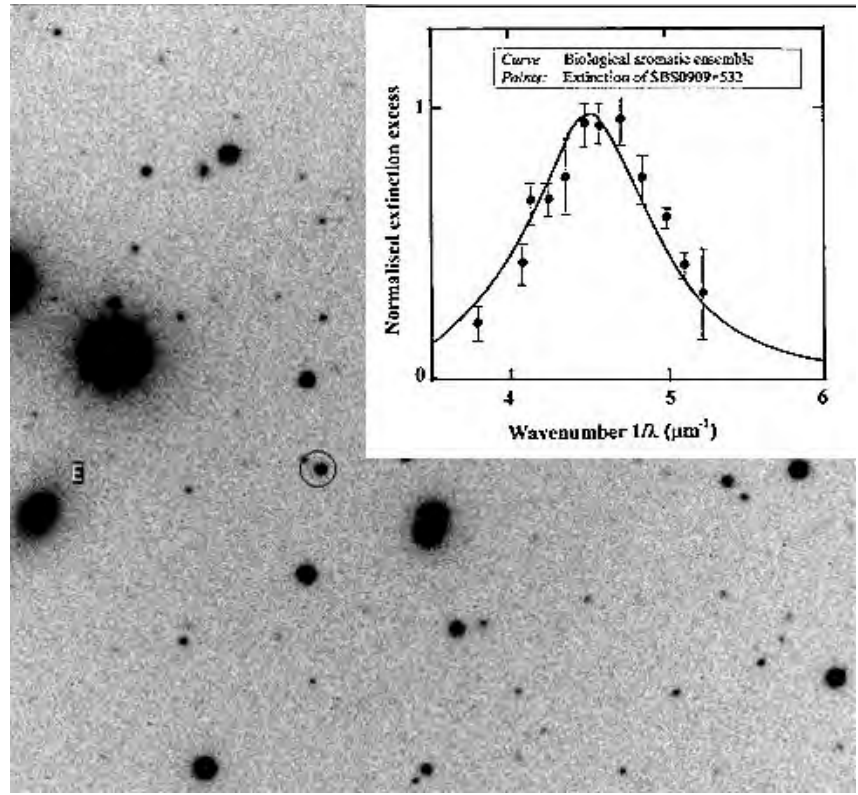

Fig.6 Agreement between the 2175A absorption of biomolecules and the data for dust in the galaxy SBS0909+532 at red shift z=0.8. This corresponds to a distance of nearly 8 billion light years – $10^{26}$ m, most of our world (See Wickramasinghe et al (2004) for details).

Although astronomers still seek abiotic models to explain the data such as in of Fig 5 and 6, biology provides by far the simplest self-consistent model. In particular, a claim that the strong peak of interstellar extinction at 2175A can be explained by abiotic aromatics (PAHs) could be seriously flawed (Hoyle and Wickramasinghe, 2000; Rauf and Wickramasinghe, 2009). Aromatic molecules resulting from the decay, degradation or combustion of biomaterial may be similar to soot or anthracite. Fig. 7 shows striking correspondences between astronomical data and such models.

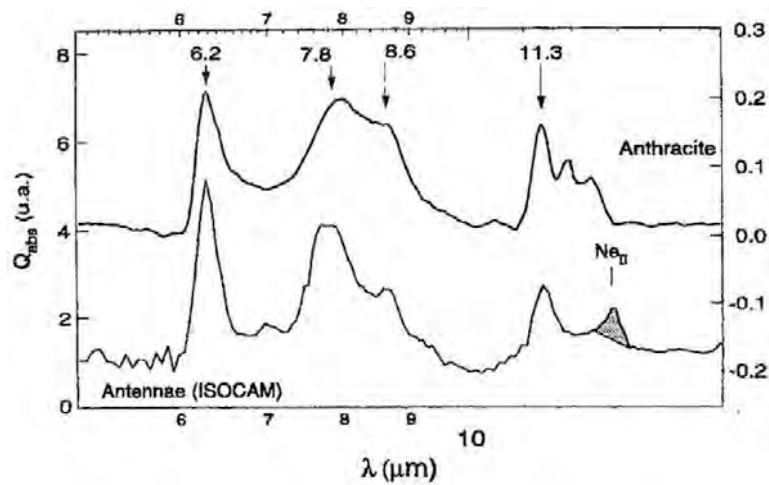

Fig. 7 Emission from dust in Antennae galaxies compared to anthracite, a biological degradation product.

## 8. Horizontal gene transfer across the galaxy

Whilst amplification of microorganisms within primordial comets could supply a steady source of primitive life (archeae and bacteria) to interstellar clouds and thence to planetary systems, the genetic products of evolved life could also be disseminated on a galaxy-wide scale. Such ideas were pioneered and developed by Joseph (2000, 2009b,c) and discussed also by others (Hoyle and Wickramasinghe, 1980, Napier, 2004, Wallis and Wickramasinghe, 2004; Wickramsinghe and Napier, 2008).

As first detailed and theorized by Joseph (2000), when prokaryotes, accompanied by viruses, were deposited onto planets already harboring life, genes were exchanged via horizontal gene transfer utilizing the same genetic mechanisms of exchange which are common among the microbes and viruses of Earth. These space-journeying microbes and viruses also exchanged and obtained genes from eukaryotes on innumerable planets. Microbes and viruses therefore began building up vast genetic libraries of genes coding for advanced and complex characteristics, and those shaped by natural selection. These genetic libraries maintained in the genomes of viruses and prokaryotes, made it possible to not only immediately adapt to every conceivable environment, but to biologically modify and terraform new planets and to promote the *evolution* and even the replication of species which had evolved on other worlds; a process Joseph (2000, 2009b,c) refers to as "evolutionary metamorphosis" and which he likens to embryogenesis. Hoyle and Wickramasinghe (2000), focusing on viruses, have referred to this as "evolution from space."

Hoyle and Wickramasinghe (2000) also pointed out, just as new computer programs can create errors in the computer's hardware, that in rare instances viruses also damage the genetic hardware of the host. However, for the most part, viruses deposited on Earth from passing comets, actually contain genes which promote evolution. When viruses from space insert their genes into the eukaryotic genome, the result is often of benefit to the host, just as most computer programs can enhance the functionality of the computer.

Joseph (2000, 2009b,c) sums it up this way: "Just as an apple seed contains the genetic instructions for the development of an apple tree, these genetic seeds of life contained the genetic instructions for the tree of life, and for every creature which has walked, crawled, swam, or slithered across the Earth...Genes act on the environment, and the biologically altered environment acts on gene selection, thereby expressing traits which had been encoded into genes acquired from life on other planets."

Therefore, according to Joseph, Hoyle and Wickramasinghe, evolution on Earth has been greatly influenced by genes acquired on other worlds. Because of the functionality of the "universal genetic code" and due to horizontal gene exchange, evolution has been coordinated on a galaxy-wide scale by microbes and viruses which have acquired and exchanged genes with microbes, viruses, and eukaryotes which evolved on innumerable planets and in every conceivable environment.

Intergalactic gene exchange is made possible by comets. Our present-day solar system appears to be surrounded by an extended halo of some 100 billion comets (the Oort Cloud) moves around the centre of the galaxy with a period of 240My. In fact, the term Oort cavity might be more appropriate, since an Oort cavity of size $r_O = (M/\varrho)^{1/3}$ forms in a PGC when planets merge to form a protostar of mass M, where ϱ is the PGC mass density.

Every 40 million years, on the average, the comet cloud becomes perturbed due to the close passage of a molecular cloud. Gravitational interaction then leads to hundreds of comets from the Oort Cloud being injected into the inner planetary system, some to collide with the Earth. Such collisions can not only cause extinctions of species (as one impact surely did 65 million years ago, resulting in the extinction of the dinosaurs), but they could also trigger the expulsion of surface material back into space. A fraction of the Earth-debris so expelled survives shock-heating and could be laden with viable microbial ecologies as well as genes of evolved life. Such life-bearing material could reach newly forming planetary systems in the passing molecular cloud within a few hundred million years of an ejection event. A new planetary system thus comes to be infected with terrestrial microbes terrestrial genes that can contribute, via horizontal gene transfer, to an ongoing process of local biological evolution.

Once life has got started and evolved on an alien planet or planets of the new system the same process can be repeated (via comet collisions) transferring genetic material carrying local evolutionary 'experience' to other molecular clouds and other nascent planetary systems. If every life-bearing planet transfers genes in this way to more than one other planetary system (say 1.1 on the average) with a characteristic time of 40My then the number of seeded planets after 9 billion years (lifetime of the galaxy) is $(1.1)^{9000/40} \sim 2 \times 10^9$. Such a large number of 'infected' planets illustrates that evolution, involving horizontal gene transfer, must operate not only on the Earth or within the confines of the solar system but on a truly galactic scale (Joseph 2000; Joseph and Wickramasinghe 2010a,b). Life throughout the galaxy on this picture would constitute a single connected biosphere.

**9. Life on other planets**

Much astrobiological attention is being focussed nowadays on the planet Mars with attempts to find evidence of contemporary life, fossil life and potential life habitats. The Jovian moon Europa, the Venusian atmosphere, the outer planets and comets are also on the astrobiologist's agenda but further down the time-line. The unambiguous discovery of life on any one of these solar system objects would be a major scientific breakthrough and would offer the first direct test of the concept of an interconnected biosphere.

The discovery of bacteria and archaea occupying the harshest environments on Earth continues to provide indirect support for panspermia. Viable transfers of microbial life from one cosmic habitat to another requires endurance of high and low temperatures as well as exposure to low fluxes of ionising radiation delivered over astronomical timescales, typically millions of years. The closest terrestrial analogue to this latter situation exists for microorganisms exposed to the natural radioactivity of the Earth, an average flux of about 1 rad per year. Quite remarkably microbial survival under such conditions is well documented. Dormant microorganisms in the guts of insects trapped in amber have been revived and cultured after 25-40 million years (Cano and Borucki, 1995); and a microbial population recovered from 8 My old ices has shown evidence of surviving DNA (Biddle et al, 2007). All this goes to show that arguments used in the past to 'disprove' panspermia on the grounds of survivability during interstellar transport are likely to be seriously flawed.

## 10. Microfossils in meteorites

The topic of microfossils in carbonaceous chondrites has sparked bitter controversy ever since it was first suggested in the mid-1960's (Claus, Nagy and Europa, 1963). Since carbonaceous chondrites are generally believed to be derived from comets, the discovery of fossilised life forms in comets would provide strong *prima facie* evidence in support of the theory of cometary panspermia. However, claims that all the micro structures (organised elements) discovered in meteorites were artifacts or contaminants led to a general rejection of the microfossil identifications. The situation remained uncertain until early in 1980 when H.D. Pflug found a similar profusion of "organised elements" in ultrathin sections prepared from the Murchison meteorite, a carbonaceous chondrite that fell in Australia on 28 September 1969 (Pflug, 1984). The method adopted by Pflug was to dissolve-out the bulk of minerals present in the thin meteorite section and examine the residue in an electron microscope. These studies made it very difficult to reject the fossil identification. More recent work by Richard Hoover (2005) and his team leaves little room for any other interpretation of these structures than that they are microbial fossils (Figs 7).

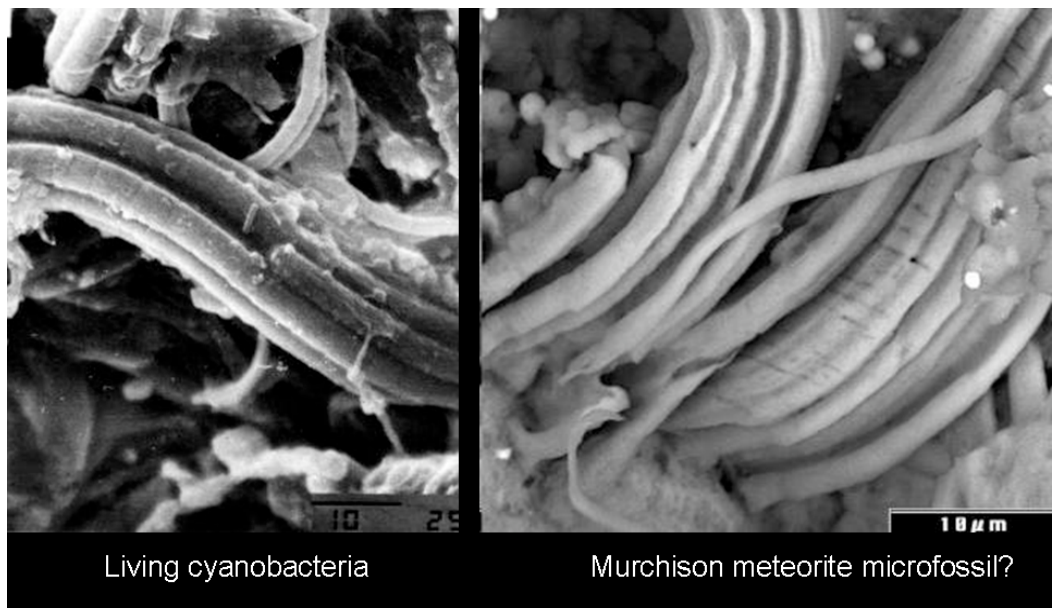

Fig.8  A structure in the Murchison meteorite compared with living cyanobacteria (Hoover, 2005)

## 11. Concluding remarks

In conclusion we note that comets are beginning to acquire a prime importance and relevance to the problem of the origin of life. It would surely be prudent to study these celestial wanderers more carefully. From 1986 onwards infrared spectra of comets have shown consistency with the presence of biologically relevant material, perhaps even intact desiccated bacteria. With some 50-100 tonnes of cometary debris entering the Earth's atmosphere on a daily basis the collection and testing of this material for signs of life should in principle at least be straightforward.  Such a project was recently started in 2001 by the Indian Space Research Organisation, ISRO, in partnership with Cardiff University. Samples of stratospheric aerosols collected using balloon-borne cryosamplers were investigated independently in Cardiff, Sheffield and India and have revealed tantalising evidence of microbial life (Harris et al, 2002; Wainwright et al, 2003, 2004). A particularly interesting component of the collected samples was in the form of 10 micrometre clumps that have were identified by SEM and fluorescence tests as being viable but not culturable microorganisms (Fig. 9).

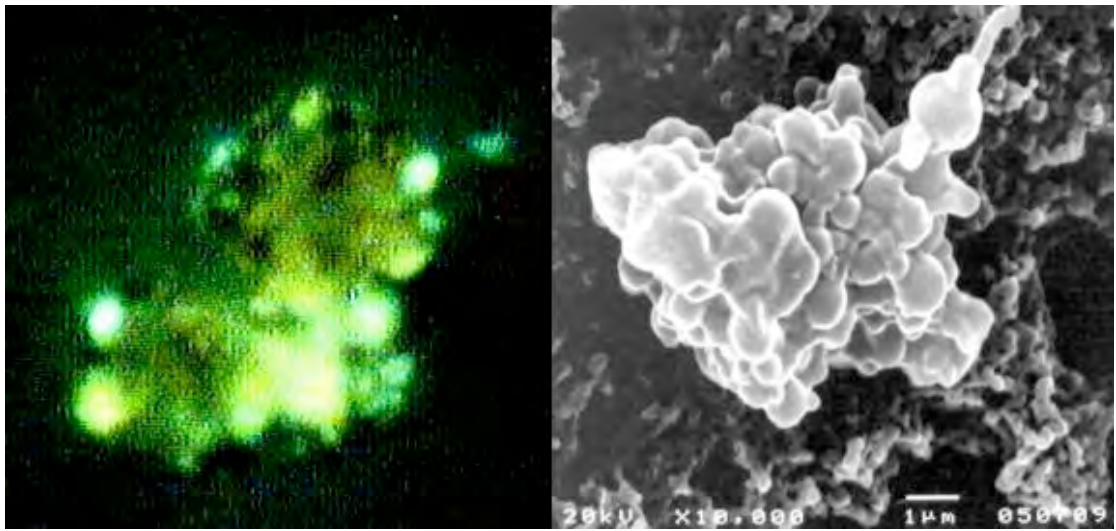

Fig. 9. Stratospheric dust collected asceptically from an altitude of 41 km showed evidence of clumps of viable but not culturable bacteria. The left panel shows a clump fluourescing under the action of a dye and the right panel shows a scanning microscope image showing a clump of cocci and a bacillus.

Because such large aggregates are virtually impossible to loft to 41 km a *prima facie* case for their extraterrestrial cometary origin has been made. However, in view of the profound importance of any conclusion such as this, it is a high priority to repeat projects of this kind. Compared with other Space Projects for solar system exploration the budgets involved are trivial, but the scientific pay-off could be huge. We might ultimately hope for confirmation that Darwinian evolution takes place not just within a closed biosphere on Earth but extends over a large and connected volume of the cosmos. This view is strongly supported by a new cosmology modified by modern fluid mechanics, where comets, moons and planet which precede all stars. The stars produced by the primordial planets then provide the fertilizer and mutational radiation, and the power for cosmic dispersion, necessary for cosmic biology.